\documentclass[aps,showpacs,preprint]{revtex4}
\usepackage{epsfig,amssymb,bm,graphics,color,amsmath}
\usepackage{graphicx}
\usepackage{slashed}
\usepackage{chngcntr}
\usepackage{float}
\restylefloat{table}

\counterwithout{footnote}{section}

\newcommand{\vsq}{v_s^2}
\newcommand{\sTt}{s/T^3}
\newcommand{\TTc}{T/T_c}
\newcommand{\dVV}{V^\prime/V}

\begin{document}{\normalsize}
\title{Cross-over versus first-order phase transition in holographic 
gravity--single-dilaton models of QCD thermodynamics}
\author{R. Yaresko, J. Knaute, B. K\"ampfer}
\address{Helmholtz-Zentrum Dresden-Rossendorf,
POB 51 01 19, 01314 Dresden, Germany and\\
TU Dresden, Institut f\"ur Theoretische Physik, 01062 Dresden, Germany}


\begin{abstract}
A dilaton potential is adjusted to recently confirmed lattice QCD thermodynamics 
data in the temperature range $(0.7 \ldots 3.5) T_c$ where $T_c = 155 \text{ MeV}$ 
is the pseudo-critical temperature. 
The employed holographic model is based on a gravity--single-field dilaton dual. 
We discuss conditions for enforcing (for the pure gluon plasma) or avoiding 
(for the QCD quark-gluon plasma) a first-order phase transition, but still 
keeping a softest point (minimum of sound velocity).
\end{abstract}

\maketitle

\section{Introduction}
The celebrated AdS/CFT correspondence \cite{AdS_CFT} has sparked a large number of 
dedicated investigations of strongly coupled systems 
(cf.\ \cite{holography_reviews} for recent surveys). 
A particular field of application is provided by the strong coupling nature of QCD 
at low momentum/energy scales. 
While employing 5d Einstein gravity in the dual description 
is strictly justified only in the 
large $N_c$ and large t'Hooft coupling limits of the boundary theory, which 
ensure suppression of loop and stringy corrections to classical gravitation theory, 
by such means 
nevertheless one could study models which are expected to exhibit a behavior resemblant of QCD. 
The aim is then often to understand, on a qualitative level, phenomena which are 
hardly accessible in the 4d quantum field theory. 
A prominent example is given by real-time phenomena, 
e.g.\ within QCD. Other phenomena, such as the hadron spectrum or the equation 
of state, are accessible by lattice QCD calculations - but here one would like to 
understand qualitatively the emerging numerical results by means of transparent models. 

Many facets of the QCD equation of state are fairly known by now, both for the physical parameter 
section and for various limits of parameters 
(e.g.\ quark masses, dimension of the gauge group, flavor number,  
adjoint representations of quarks etc.). This statement applies only for finite-temperature ($T$)
QCD at zero baryo-chemical potential ($\mu$). 
However, in relativistic 
heavy ion collisions, the bulk of excited matter has $\mu > 0$, as inferred from the 
analysis of hadron abundancies \cite{hadron_abundancies}. The knowledge of the QCD 
equation of state is, therefore, presently incomplete 
(in particular beyond the range accessible by the $\mu/T \ll 1$ expansion) 
and calls urgently for an 
improvement. In particular, there are several ideas that QCD allows for 
a critical point in the $T-\mu$ phase diagram where the cross-over turns in a 
first-order phase transition. Mainly based on universality arguments, a 
multitude of models have been employed to locate the critical point 
\cite{CBM_book, Stephanov_CP}, but also more directly QCD anchored approaches, 
e.g.\ Dyson-Schwinger equations as integral formulation of QCD, have been 
used \cite{Fischer_DSE}. Parallel to the theoretical attempts, also special 
experimental searches are conducted, e.g.\ the beam energy scan at RHIC \cite{RHIC_BES}.

Coming back to options for modeling a phase diagram similar to QCD with 
the conjectured critical point, we mention \cite{Gubser_quarks}, where in a 
holographic model, including gravity, a dilaton field and a U(1) gauge field,
the possibility of such a realisation has been demonstrated. The set-up of 
\cite{Gubser_quarks} is based on a dilation potential, 
which features qualitatively the equation of 
state at $\mu = 0$, supplemented by a dynamical strength function, 
which is adjusted to the quark number susceptibility, again at $\mu =0$. 
While in the infinitely heavy quark mass ($m_q \rightarrow \infty$) limit of QCD, 
which becomes then a pure Yang-Mills theory, the equation of state is known 
since some time \cite{Boyd} and has been confirmed by high-precision lattice QCD 
simulations \cite{Fodor_glue}, the status of QCD with physical quark masses has been
settled only very recently. After refinements in the lattice discretization schemes 
and actions and continuum extrapolations the results of two independent 
collaborations \cite{WuBp_quarks, hotQCD} became consistent. 

Given this new situation and having in mind e.g.\ an application in the spirit of \cite{Gubser_quarks} 
to the QCD phase diagram modeling, one should seek for an appropriate 
dilaton potential, reproducing sufficiently accurately the by now known QCD 
equation of state at $\mu = 0$. This is the aim of the present note. We have hereby 
the attitude to take the AdS/CFT dictionary literally, i.e.\ translate, 
without corrections due to $N_c = 3$ or finite coupling, the 5d Riemann 
metric into 4d energy-momentum tensor components (or correlators) and 
vice versa.

\section{Adjusting a dilaton potential}
At $\mu = 0$, the equation of state, in parametric form, follows from \cite{Gubser}
\begin{eqnarray}
LT(\phi_H) &=& \frac{V(\phi_H)}{\pi V(\phi_0)} \exp \Big(A(\phi_0) + \int_{\phi_0}^{\phi_H} d\phi  \Big[ \frac{1}{4X} + \frac23 X \Big]\Big), \\
G_5s(\phi_H) &=& \frac{1}{4} \exp \Big(3A(\phi_0) + \frac34\int_{\phi_0}^{\phi_H} d\phi \frac{1}{X}\Big),
\end{eqnarray}
for entropy density $s$ and temperature $T$, where the scalar function $X(\phi;\phi_H)$ \cite{Kiritsis_long}
is determined by the system (a prime means a derivative w.r.t.\ $\phi$)
\begin{eqnarray}
X^\prime &=& - \Big( 1 + Y - \frac 2 3 X^2 \Big) \Big(1 + \frac{3}{4X} \frac{V^\prime}{V}\Big), \label{eq:Xeq} \\
Y^\prime &=& - \Big( 1 + Y - \frac 2 3 X^2 \Big) \frac{Y}{X}, \label{eq:Yeq}
\end{eqnarray}
which is integrated from the horizon $\phi_H - \epsilon$ to the boundary $\phi_0$ 
with initial conditions 
\begin{eqnarray}
 X(\phi_H - \epsilon) &=& - \frac 34 \frac{V^\prime(\phi_H)}{V(\phi_H)} + \mathcal{O}(\epsilon^1), \\
 Y(\phi_H - \epsilon) &=& - \frac{X(\phi_H - \epsilon)}{\epsilon} + \mathcal{O}(\epsilon^0),
 \end{eqnarray}
and $\epsilon \rightarrow 0$.
The quantity $A(\phi_0)$ encodes the near-boundary behavior of the model. 
We assume $L^2 V(\phi) \approx - 12 + \frac{L^2 M^2}{2}\phi^2$ for $\phi \rightarrow \phi_0 = 0$ 
which results in $A(\phi_0) = \frac{\log \phi_0}{\Delta - 4}$, whereby 
we have set $L\Lambda = 1$ \cite{Gubser} and, as usual, $L^2 M^2 = \Delta(\Delta - 4)$. 
We consider $2 < \Delta < 4$. 

From $s$ and $T$, the pressure follows as 
\begin{equation}
 p(\phi_H) = \int_{\infty}^{\phi_H} d\tilde\phi_H \frac{dT(\tilde\phi_H)}{d\tilde \phi_H} s(\tilde\phi_H), \label{eq:p}
\end{equation}
where $p(\infty) = 0$ holds if $\dVV < 2 \sqrt{2/3}$ with 
$\dVV\vert_{\phi \rightarrow \infty} \rightarrow \text{const}$, 
corresponding to a "good" singularity at $\phi = \infty$ \cite{Kiritsis_long}. 
We consider only such cases.

Besides of a proper adjustment of the dilaton potential $V(\phi)$ to the equation of state, 
the model parameters $G_5/L^3$ and $L$ must be fitted, too. 
Since a direct mapping procedure of an input equation of state to the potential 
is not at our disposal, we use as trial ansatz 
\begin{equation} \label{eq:v_D}
 v_D(\phi) \equiv \frac{V_D^\prime}{V_D}=
 \begin{cases}
   \frac{- L^2 M^2}{12}\phi + s_1 \phi^3 &\text{ for } \phi \leq \phi_m,  \\
   \Big(t_1 \tanh(t_2 \phi - t_3) + t_4\Big)\Big(1 - \frac{b_1}{\cosh(b_2\phi - b_3)^2}\Big) &\text{ for } \phi \geq \phi_m,
 \end{cases}
\end{equation}
(demanding differentiability of $v_D$ at $\phi_m$ fixes $L^2 M^2$ and $s_1$) and find 
\begin{equation} \label{eq:table}
\begin{tabular}{c|c|c|c|c|c|c|c|c|c}
 fit to   & $\phi_{m}$ & $t_{1}$ & $t_{2}$ & $t_{3}$ & $t_{4}$ & $b_{1}$ & $b_{2}$ & $b_{3}$ & $G_5/L^3$ \\ \hline
  $v_s^2$  & 0.2163 & 0.6453 & 0.4988 & 0.0845 & 0.0286 & 0.4842 & 2.5020 & 3.9887 & 0.4544 \\ \hline
  $s/T^3$  & 0.2430 & 0.6480 & 0.5023 & 0.0855 & 0.0344 & 0.4844 & 2.6162 & 4.1458 & 0.4586 
\end{tabular} 
\end{equation}
with $LT_c = 1.8036$ (fit to $\vsq$ from \cite{WuBp_quarks}) or $LT_c = 0.5051$ (fit to $\sTt$ from \cite{WuBp_quarks}).
This ansatz obeys the 
Chamblin-Reall IR behavior $L^2 V(\phi \to \infty) \sim e^{(t_1 + t_4) \phi}$. 
The approach belongs to a similar class of holographic models as the model 
class B in \cite{Noronha}: it has no confinement in 
the sense of \cite{Kiritsis_long} for $t_1 + t_4 < \sqrt{2/3}$ and no explicit 
fermionic degrees of freedom. Our ansatz is meant purely to match lattice QCD thermodynamics data 
in a restricted temperature interval. It is therefore interesting to see in future 
investigations, e.g.\ whether the explicit account of quarks has a similar impact 
on $\zeta/s$ as found in the present setting. 
To set a scale, we determine $T_c$ in the holographic model by the inflection 
point of $s/T^3$ as a function of $T$, and $T_c = 155 \text{ MeV}$ \cite{WuBp_quarks} is used 
in the lattice QCD data \cite{WuBp_quarks}. The resulting velocity of sound squared, 
$v_s^2 =\frac{d \log T}{d \log s} = \frac{d\log T}{d \phi_H}\big(\frac{d\log s}{d \phi_H} \big)^{-1}$, the scaled entropy density, $s/T^3$, the 
scaled pressure, $p/T^4$, and the scaled interaction measure 
$I/T^4 = (sT - 4p)/T^4$ are exhibited in Fig.~1 together with the lattice QCD 
data \cite{WuBp_quarks, hotQCD}. The solid blue (dotted red) curves are our optimum fits of 
$v_s^2$ ($s/T^3$) 
with the parameters of \eqref{eq:table}. Circles depict the respective quantities at $T_c$. One 
observes that the softest point, i.e.\ the minimum of $\vsq$ as a function of $\TTc$, 
is slightly below unity (see upper left panel in Fig.~1) and the maximum of the interaction 
measure is a little bit up-shifted in comparison with the lattice QCD results 
(see lower right panel in Fig.~1). One observes also some other minor imperfections of our 
fits, in particular at the lowest temperatures covered by the lattice QCD data, 
and also at large temperatures for the interaction measure. 

\begin{figure}[!h]
\begin{center}
\includegraphics[width=0.495\textwidth]{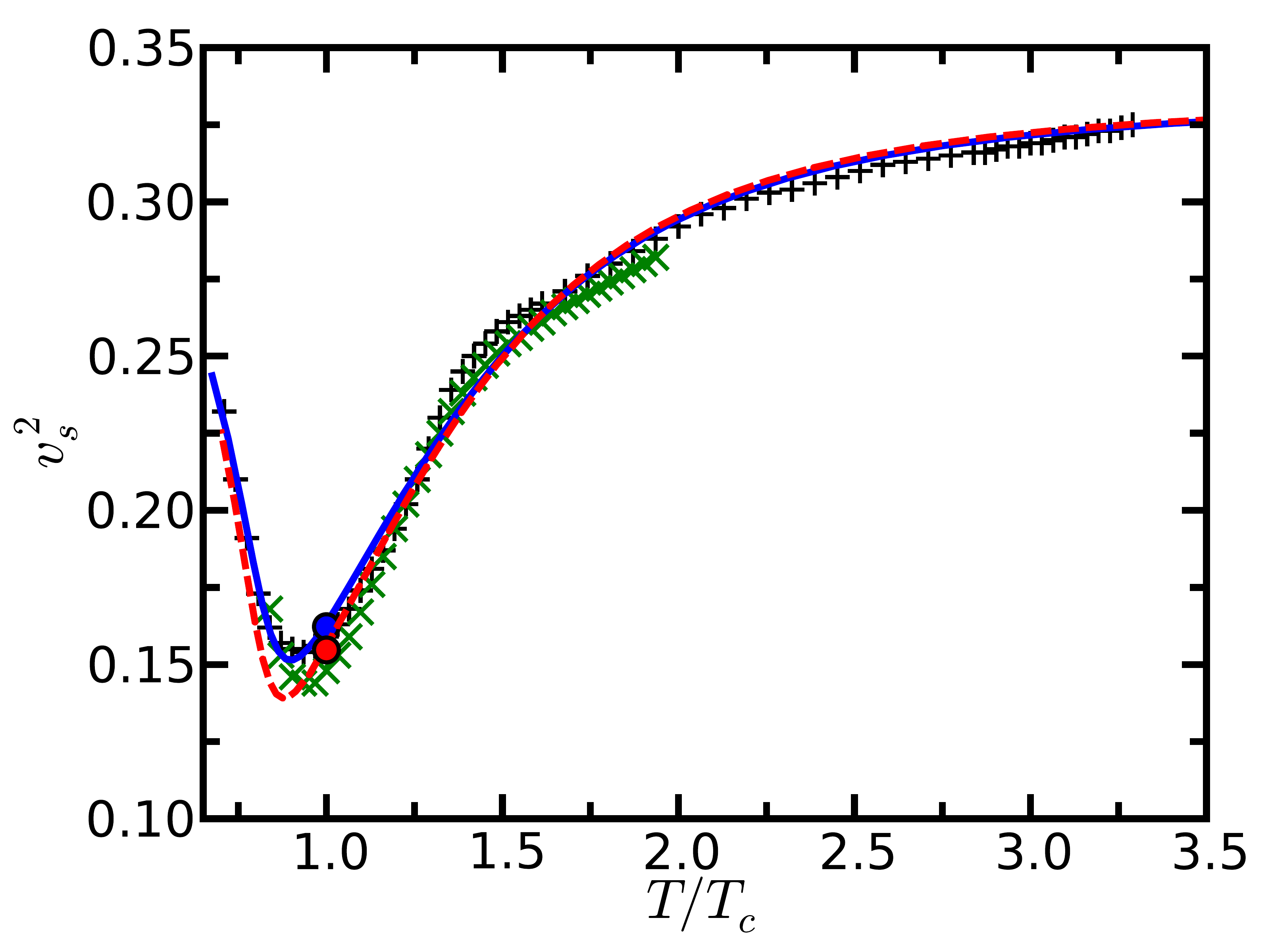}
\includegraphics[width=0.495\textwidth]{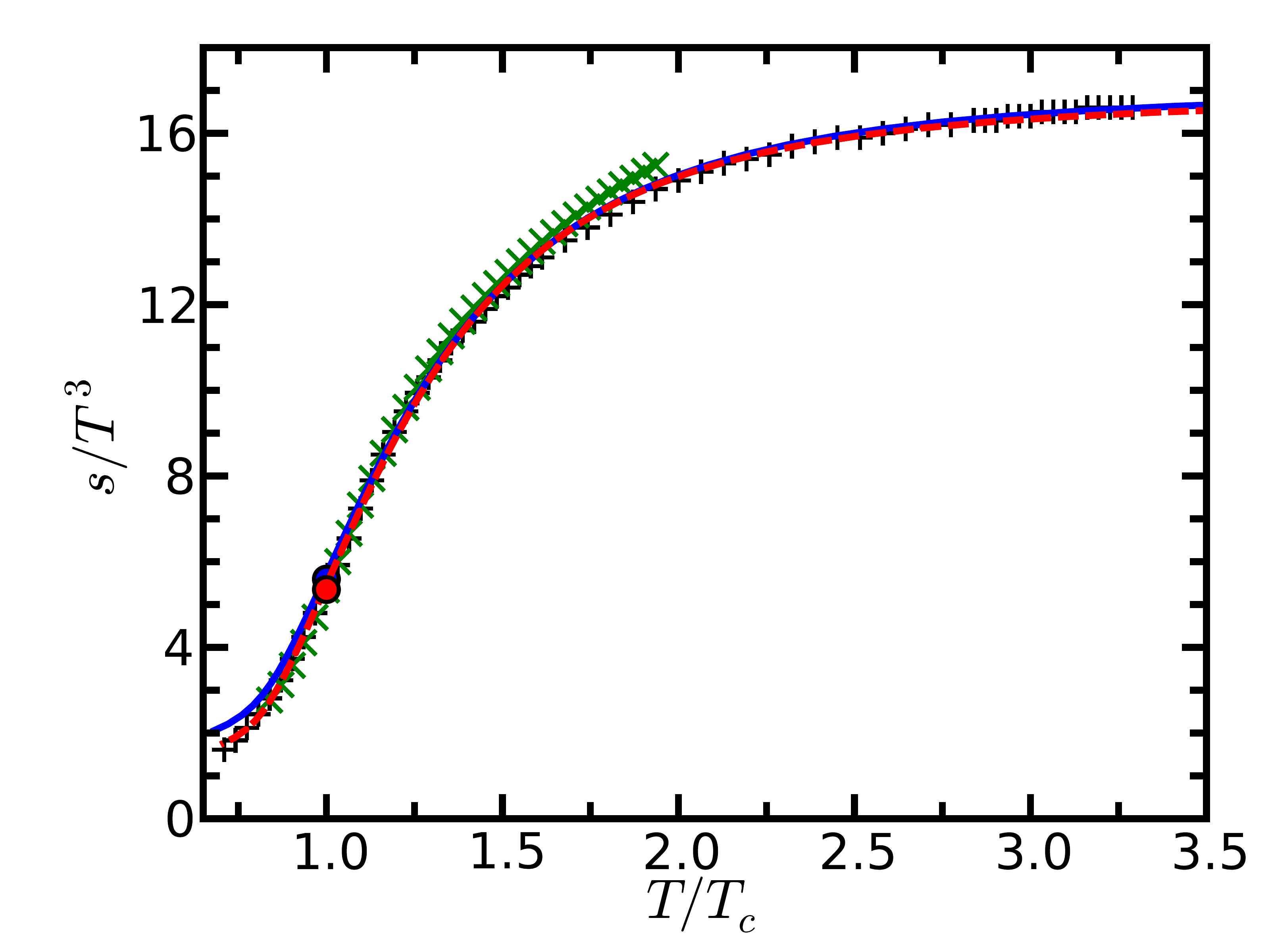}
\includegraphics[width=0.495\textwidth]{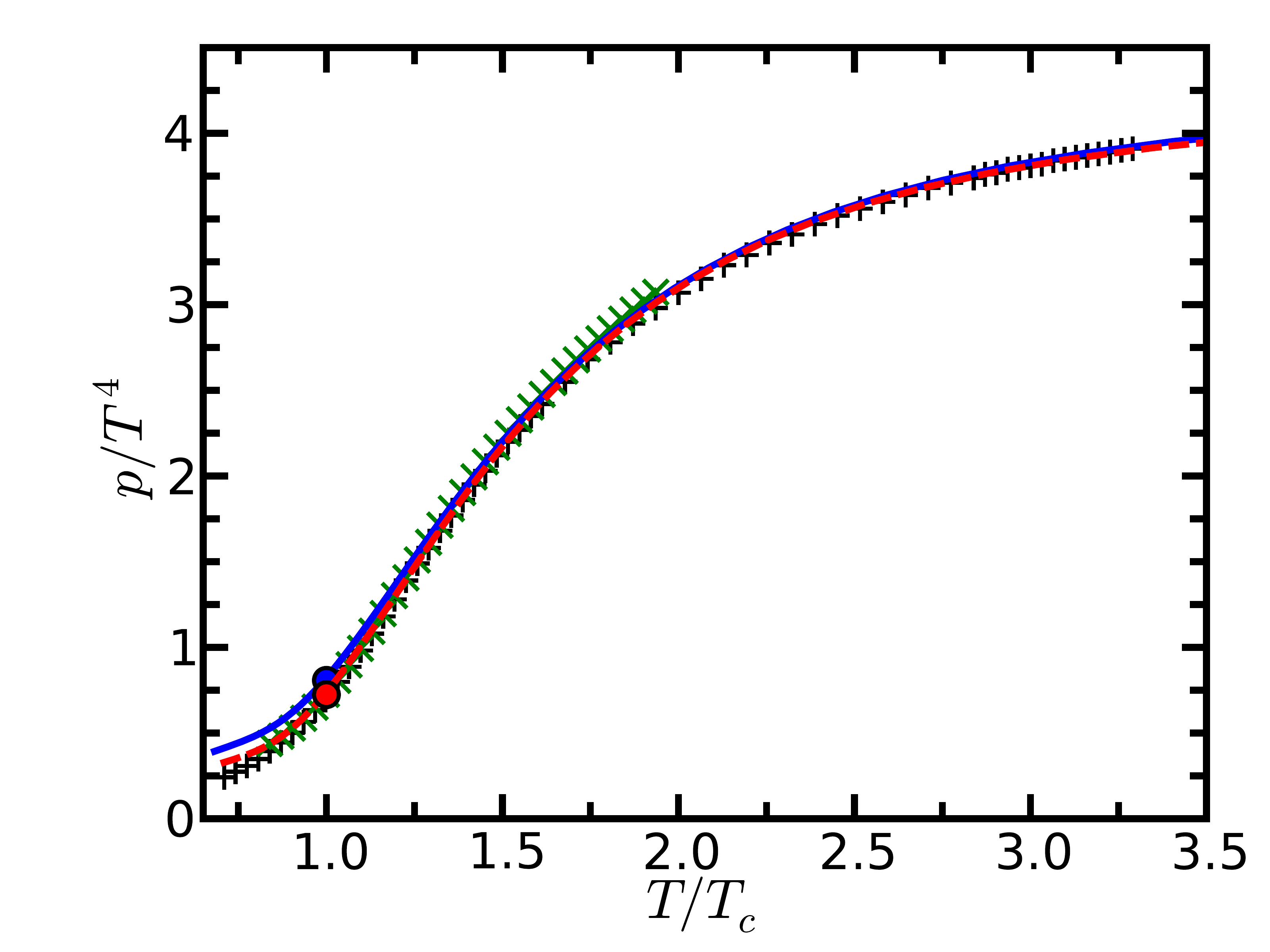}
\includegraphics[width=0.495\textwidth]{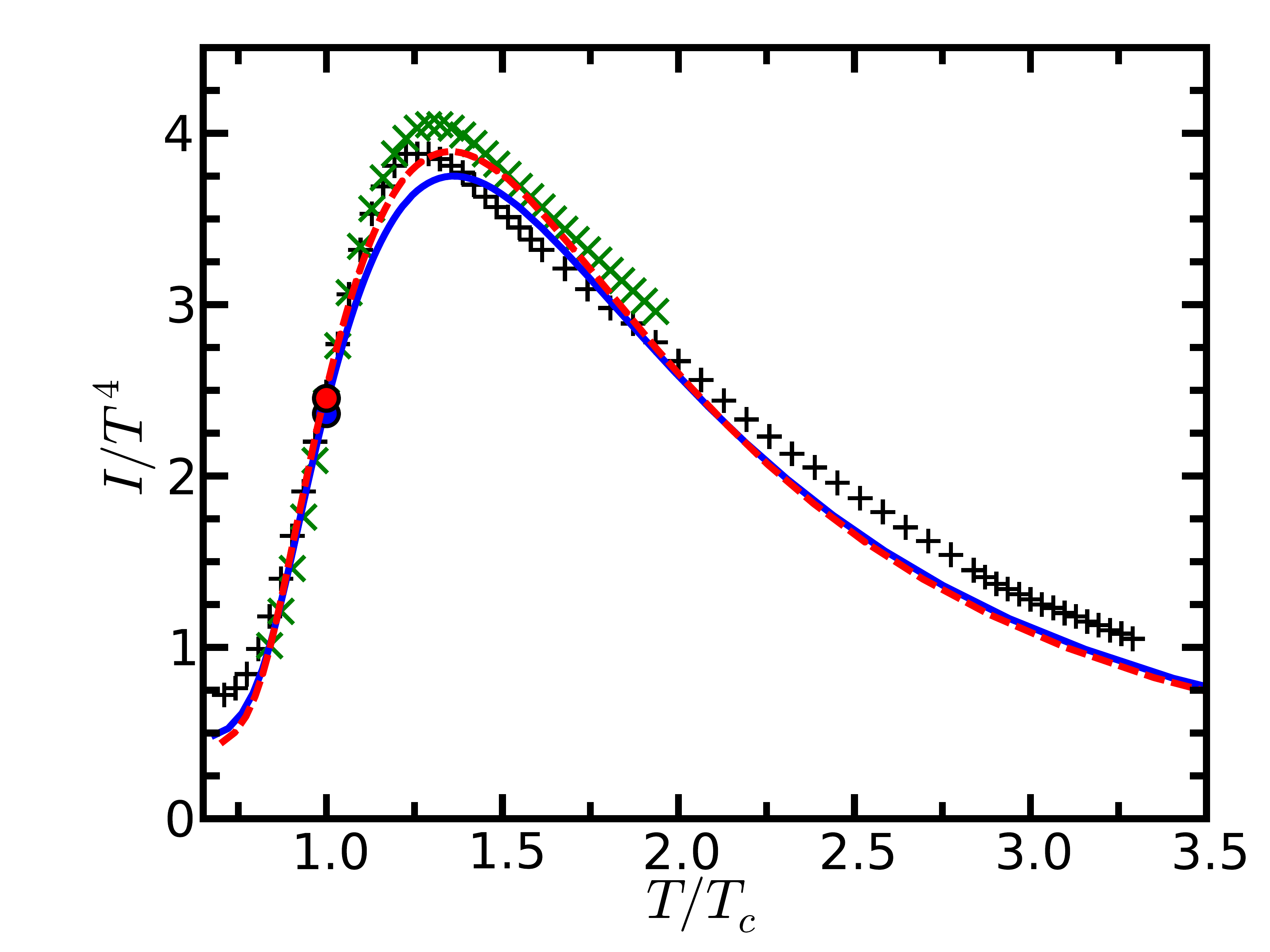}
\end{center}
\caption{
Velocity of sound squared $v_s^2$ (upper left panel),
scaled entropy density $s/T^3$ (upper right panel),
scaled pressure $p/T^4$ (lower left panel), and
scaled interaction measure $I/T^4$ (lower right panel)
as functions of $T/T_c$. 
Solid blue curves: fit to $\vsq$ data, dashed red curves: fit to $\sTt$ data, 
circles: position of thermodynamic quantities at $T_c$. Lattice QCD data: 
black plusses from \cite{WuBp_quarks}, green crosses from \cite{hotQCD};
for both sets we use the pseudo-critical temperature $T_c = 155 \text{ MeV}$.
\label{fig.1}}
\end{figure}

In the present setting, the ratio of shear viscosity to entropy density is 
$\eta/s = 1/(4\pi)$ \cite{KSS}, as usual for the Hilbert action on the gravity side \cite{GPR,EO}. 
The ratio of bulk to shear viscosity can be calculated via the Eling-Oz formula \cite{EO}
\begin{equation}
 \frac{\zeta}{\eta}\Big\vert_{\phi_H} = \Big( \frac{d \log s}{d \phi_H} \Big)^{-2} = 
 \Big( \frac{1}{v_s^2}\frac{d \log T}{d \phi_H} \Big)^{-2}, \label{eq:zeta_EO}
\end{equation}
or, equivalently \cite{EO_Buchel_Kiritsis}, via the Gubser-Pufu-Rocha formula \cite{GPR}
\begin{equation}
 \frac{\zeta}{\eta}\Big\vert_{\phi_H} = \frac{V^\prime(\phi_H)^2}{V(\phi_H)^2}
\frac{1}{\vert h_{11}(\phi_0) \vert^2}, \label{eq:zeta_GPR}
\end{equation}
where $h_{11}(\phi_0)$ is extracted from solving the perturbation equation
\begin{equation}
 h_{11}^{\prime \prime} + 
 \frac{1}{X} \Big( 1 + Y - \frac 2 3 X^2 \Big) \Big( 2 + \frac{9}{4X} \frac{V^\prime}{V}\Big) h_{11}^\prime - 
 \frac{Y}{X^2} \Big( 1 + Y - \frac 2 3 X^2 \Big) \Big(1 + \frac{3}{4 X} \frac{V^\prime}{V}\Big) h_{11} = 0, 
 \label{eq:h11_eq}
\end{equation}
with initial conditions $h_{11} (\phi_H - \epsilon) = 1$ and $h_{11}' (\phi_H - \epsilon) = 0$
for $\epsilon \to 0$.
\begin{figure}[!h]
\begin{center}
 \includegraphics[width=0.48\textwidth]{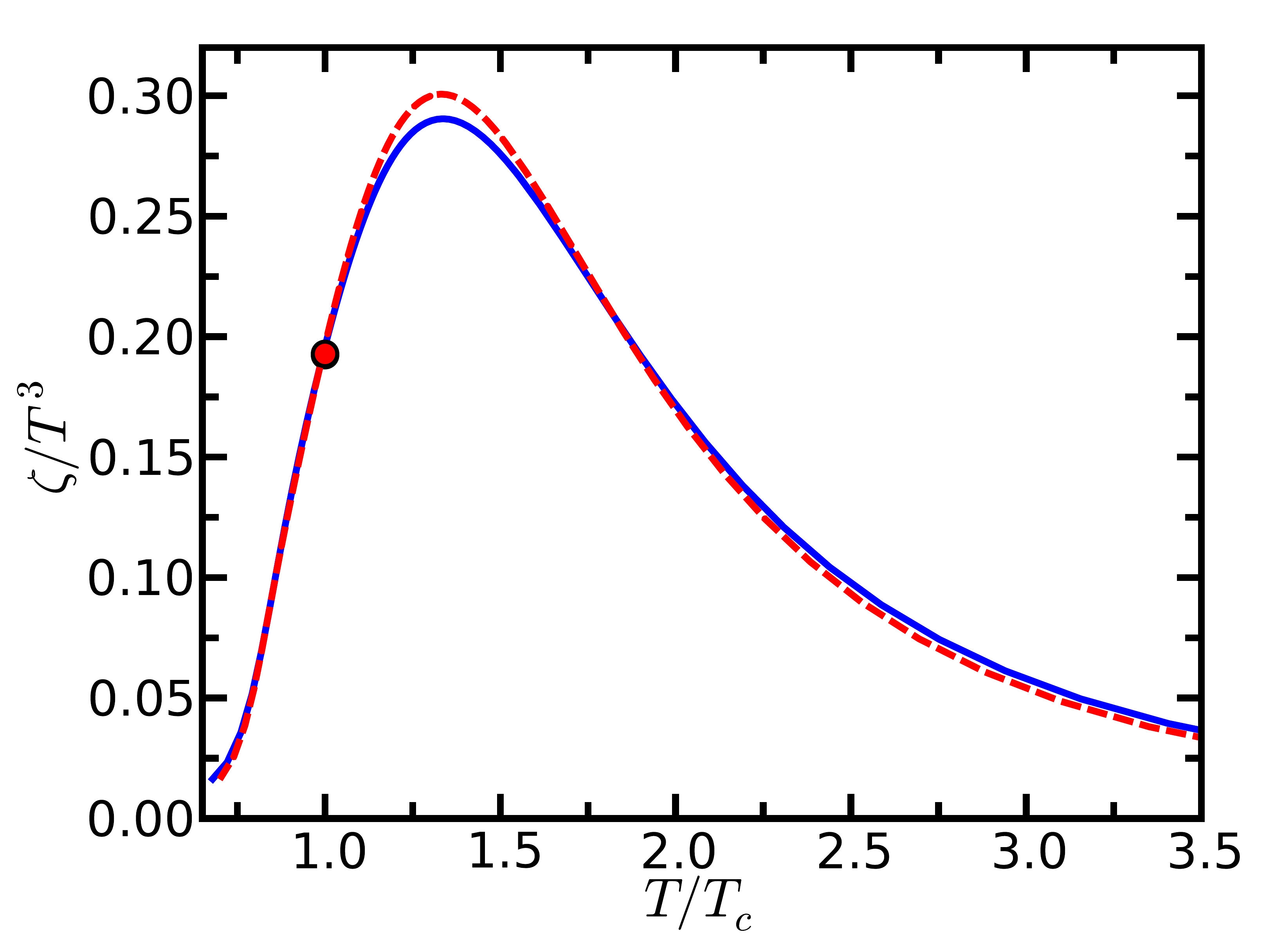}
  \includegraphics[width=0.48\textwidth]{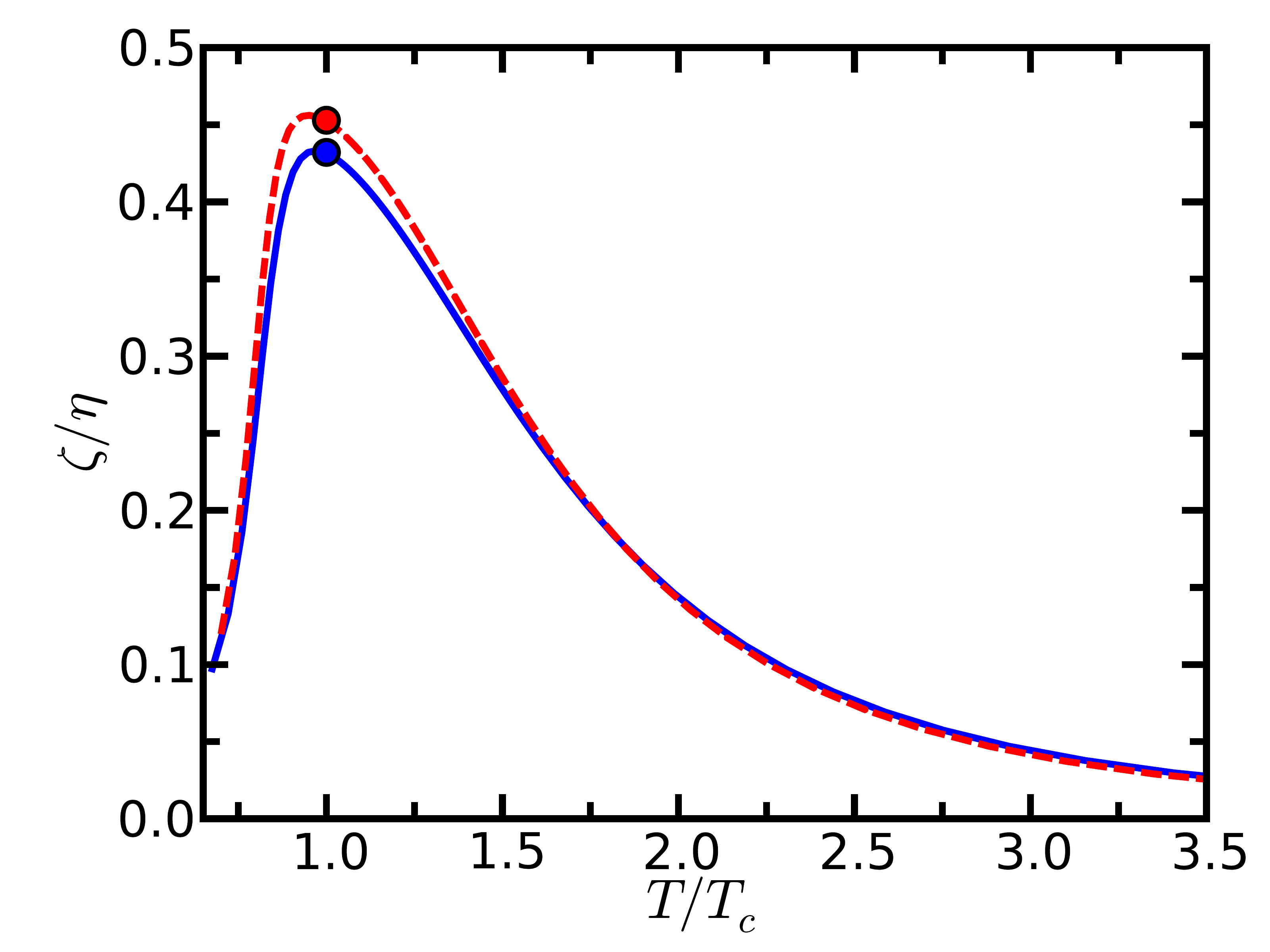}
\end{center}
\caption{Scaled bulk viscosity, $\zeta/T^3$, (left) and 
ratio of bulk to shear viscosity, $\zeta/\eta$, (right) as a function 
of $\TTc$. Line codes as in Fig.~1.
\label{fig.2}}
\end{figure}
The result is exhibited in Fig.~2. Remarkable is the reduction of $\zeta/\eta$ by 
$50\%$ at $T_c$ in comparison with the SU(3) gluon plasma (YM) considered in \cite{RYBK}. 
To understand this difference, recall the adiabatic approximation 
of \cite{Gubser}: $X(\phi) \approx - \frac 34 \frac{V^\prime(\phi)}{V(\phi)}$. 
In this approximation, the non-local term in \eqref{eq:zeta_GPR} becomes unity, 
$h_{11} = 1$ 
(since the coefficient of $h_{11}$ in \eqref{eq:h11_eq} vanishes, 
see also \cite{Kiritsis_zeta}), 
and, comparing the values of $V^\prime/V \approx 0.6 \, (0.8)$
for the QGP (pure glue, see below) at $T_c$ we find the ratio 
$\big((\zeta/\eta)^{QGP}/(\zeta/\eta)^{YM}\big)(T_c) \approx 56\%$ 
(cf.\ also \cite{Mei,Noronha} for recent holographic calculations of transport coefficients). 

On the other hand, for $\zeta/T^3$, the situation is reversed: at $T_c$, the QGP value 
is $50\%$ larger in comparison to the gluon plasma case; the peak of $\zeta/T^3$ is located at 
a larger value $T/T_c \approx 1.3$. This difference between $\zeta/\eta$ and $\zeta/T^3$ 
for QGP and for the gluon plasma can be attributed to the different number of 
degrees of freedom as reflected by the scaled entropy density $s/T^3$.

\section{Cross-over vs.\ first-order phase transition}
Remarkably, the ansatz \eqref{eq:v_D} for the dilaton potential is the same as used in \cite{RYBK} 
to describe the SU(3) gluon plasma, which displays a first-order phase transition. 
(This is actually not so surprising, as \cite{Gubser} has demonstrated that a 
two-parameter ansatz for the potential allows either for a cross-over, 
or a first-order phase transition or a second-order transition, 
depending on the choice of the parameters. Also, 
\cite{Noronha} uses a unique ansatz with two parameter sets to arrive at 
a first-order phase transition or a cross-over.) 
To elucidate the origin of such a difference we exhibit in Fig.~3 a few 
relevant quantities of both optimized models. 
\begin{figure}[!h]
\begin{center}
\includegraphics[width=0.495\textwidth]{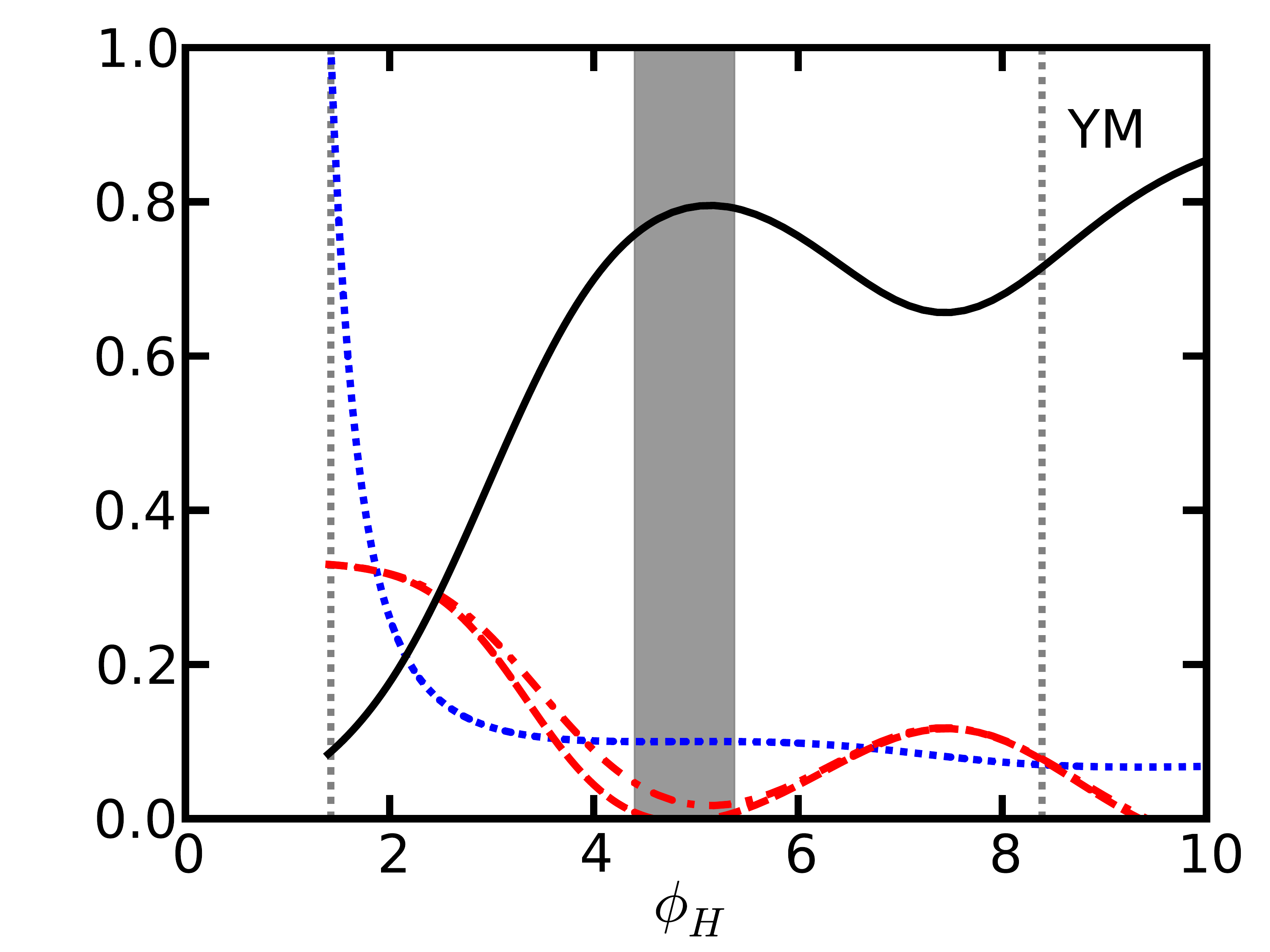}
\includegraphics[width=0.495\textwidth]{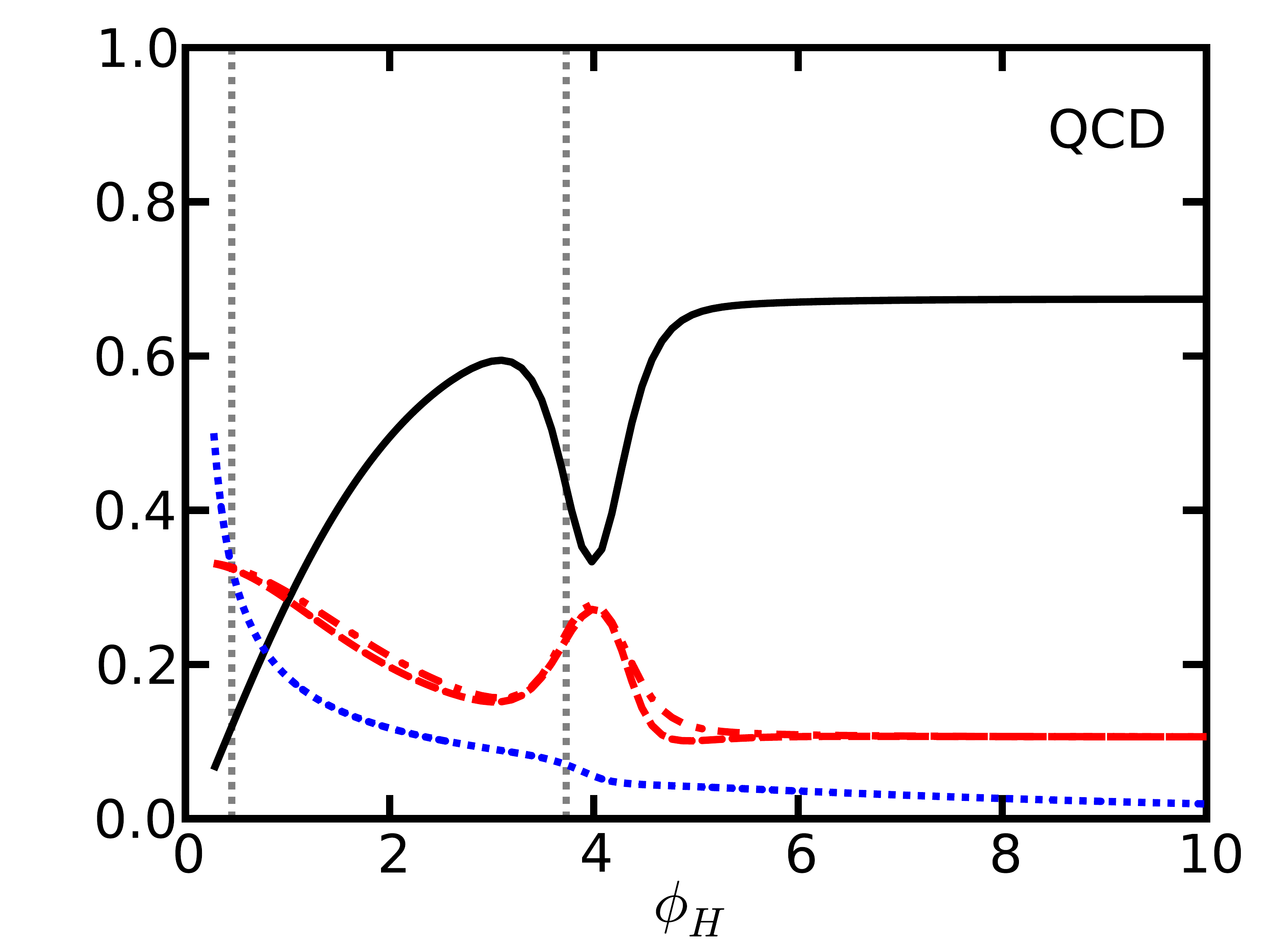}
\end{center}
\caption{$\dVV$ (solid black curves), $\vsq$ (dashed red curves, dot-dashed 
curves are for the adiabatic approximation) and $0.1\TTc$ (blue dotted curves) 
as functions of $\phi_H$. Left panel: for the pure gluon plasma (the 
grey band covers the unstable and metastable regions), right panel: 
for QCD quark-gluon plasma (fit to $\vsq$). Vertical dotted lines bracket the fit range to the 
lattice data.
\label{fig.3}}
\end{figure}

For the pure gluon plasma 
(left panel) the quantity $V^\prime/V$ has a first maximum of about $0.8$ at 
$\phi_H \approx 5.1$. In the adiabatic approximation \cite{Gubser} the 
velocity of sound squared is 
\begin{equation} \label{eq:vsq_AA}
\vsq \approx \frac 13 - \frac 12 \Big(\frac{V^\prime}{V} \Big)^2 + \ldots ,
\end{equation}
i.e.\ a local maximum (minimum) of $\dVV$ is related to a local minimum (maximum) of $\vsq$. 
If $\dVV$ is 
sufficiently large, $\vsq$ can go to zero. In fact, the adiabatic approximation 
is quite accurate (compare the red dot-dashed curve (adiabatic approximation) and 
the dashed curve (exact result) in Fig.~3 - left panel). That implies, lifting $\dVV$ 
sufficiently causes a first-order phase transition, here signalled by $\vsq = 0$. 
The entropy density $s(\phi_H)$ is a monotonously dropping function, 
as holds true for the considered examples (and is assumed to hold in general 
in the thermodynamically stable phase, see \cite{Kiritsis_long}).
Hence, $\vsq = 0$ corresponds to an extremum of $T(\phi_H)$, which is a 
minimum (maximum) if $d\vsq/d\phi_H < 0$ ($d\vsq/d\phi_H > 0$). 
Thus, if $\dVV$ is adjusted such that $\vsq(\phi_H)$ becomes negative 
in some $\phi_H$ interval and then, for larger $\phi_H$, rises to become positive again, 
the local minimum of $T(\phi_H)$ is followed by a local 
maximum. (These extrema are very shallow in the left 
panel of Fig.~3 and hardly visible on the used scale.) Such a behavior of $T(\phi_H)$ 
leads in turn to the usual loop structure of $p(T)$, characteristic of a first-order 
phase transition.
For the case at hand, the pressure is always positive. 
(In contrast, the IHQCD model \cite{Kiritsis_long} has one 
global minimum of $T(\phi_H)$ which gives rise to the high-temperature 
branch and the unstable section of $p(\phi_H)$; the low-temperature branch 
is represented by the line $p = 0$ corresponding to the thermal gas.)

Inspection of the same quantities for our fit of the QCD equation of state 
(see right panel in Fig.~3) reveals $\dVV < 0.8$ everywhere in the considered 
range of $\phi$ from $0$ (UV region) up to $10$ (towards the IR region), 
that is $\vsq > 0$ everywhere. Also here, the adiabatic approximation is 
quite accurate (compare the red dot-dashed curve (adiabatic approximation) and 
the dashed curve (exact result) in Fig.~3 - right panel). 

Note that Fig.~3 also unravels some uncomfortable 
features of the ansatz \eqref{eq:v_D} with parameters adjusted to the 
lattice Yang-Mills equation of state as in \cite{RYBK}: To catch the shape of 
thermodynamic quantities in the temperature range $(0.7 - 10)T_c$, 
the ansatz \eqref{eq:v_D} does not qualify to continue towards the deep IR 
region, since, e.g., $\vsq$ becomes negative for $\phi_H \gtrsim 9.5$, 
signalling the break-down of the ansatz's capabilities. 
(From the IHQCD viewpoint such a behavior is admissible: 
the point where $\vsq = 0$ would signal a Hawking-Page phase transition to 
the $p = 0$ phase, and desirable: the model becomes zero-$T$ confining \cite{Kiritsis_long}. 
In contrast, our ansatz \eqref{eq:v_D} is an ad hoc construction to mimic the 
Yang-Mills equation of state for $T > 0.7T_c$ (up to $10 T_c$), corresponding to 
$\phi_H \lesssim 8.5$ (down to $\phi_H \approx 1.5$). It can be supplemented by further 
terms becoming relevant for $\phi_H \gtrsim 8.5$. Thus, it is meaningless to derive 
from \eqref{eq:v_D} properties of the boundary theory in the IR region.)
In contrast, for the QCD parameter adjustment (see \eqref{eq:table}), 
the ansatz \eqref{eq:v_D} seems to be applicable towards the deep IR region. 

Upon an integration of $\dVV$ the potentials $V(\phi)$ emerge,  
displayed in Fig.~4. In contrast to $\dVV$, the potentials look quite featureless, 
both in the region where the softest point (minimum of $\vsq$) appears 
for the QCD equation of state (depicted by the arrows) and in the region of the 
first-order phase transition for the Yang-Mills equation of state (grey polygon).

\begin{figure}[!ht]
\begin{center}
\includegraphics[width=0.7\textwidth]{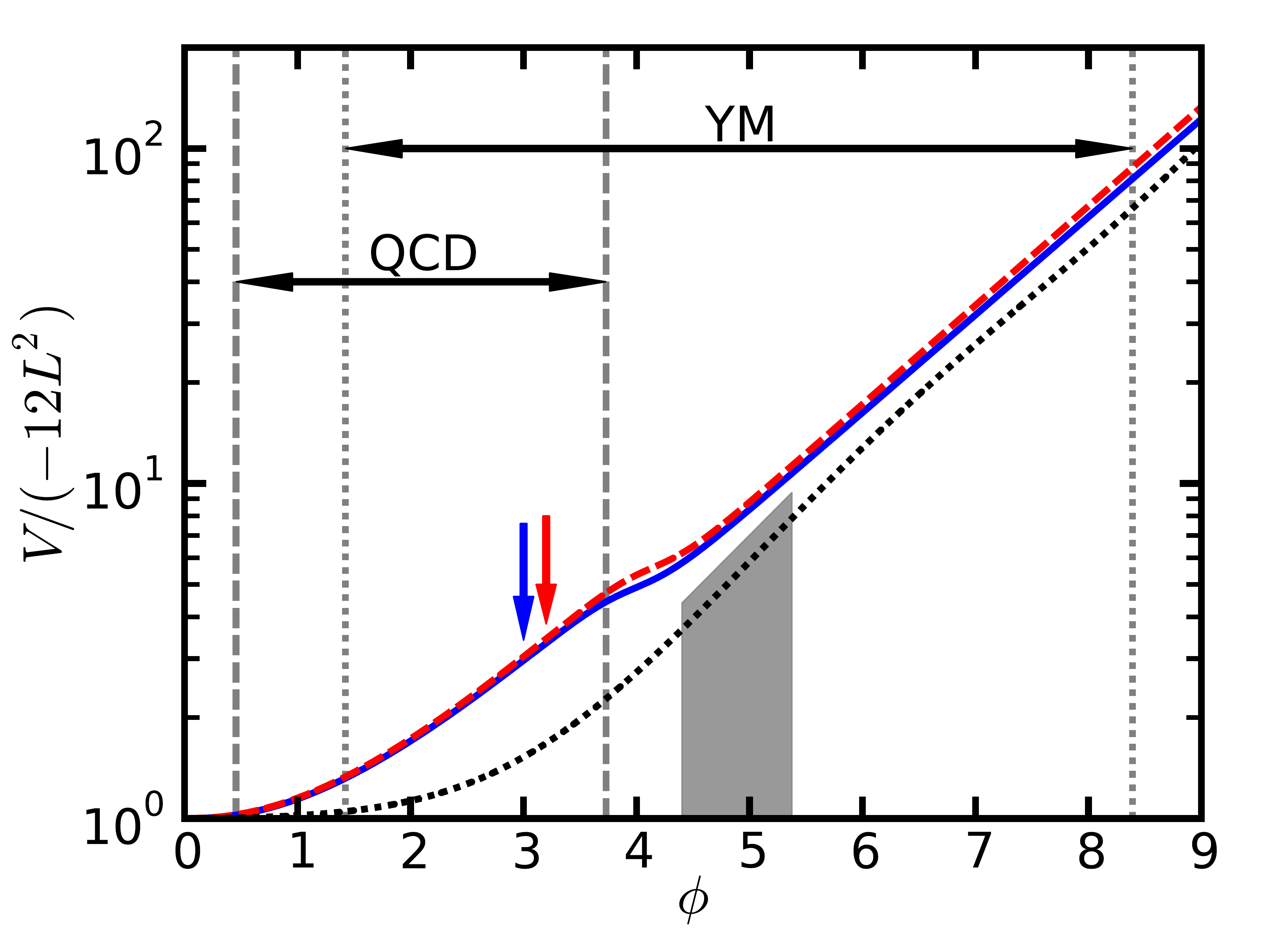}
\end{center}
\caption{The dilaton potentials $V(\phi)$ in units of $-12L^2$ for the ansatz \eqref{eq:v_D} with 
parameter sets from \eqref{eq:table} (solid blue or red dashed curve for the fit of $\vsq$ or $\sTt$; 
the arrows point to the location where $\vsq$ has the minimum). 
The potential \eqref{eq:v_D} with parameters adjusted to the Yang-Mills equation 
of state \cite{RYBK} is exhibited by the dotted black curve (the un/metastable region is 
depicted by the grey polygon). 
Vertical dashed (dotted) lines bracket the fit range to QCD \cite{WuBp_quarks} 
(Yang-Mills \cite{Fodor_glue}) lattice data.
\label{fig.4}}
\end{figure}

\section{Discussion and Summary}
In contrast to the IHQCD model, which covers quite a lot of QCD features 
both for the pure gluon plasma \cite{Kiritsis_long, Kiritsis_rev}
and for QCD in the Veneziano limit \cite{Kiritsis_VQCD}, at finite as well as  
at zero temperature together with a direct account of the two-loop t'Hooft 
running coupling, we consider here a simple holographic gravity--single-dilaton 
model without any explicit a priori scale setting. All parameters are adjusted 
to finite-temperature lattice QCD thermodynamics 
in a selected temperature range. We formulate a simple 
criterion to see already at the dilaton potential (actually its scaled derivative) 
whether a first-order phase 
transition can emerge, as for the pure gluon plasma, or a cross-over is encoded, 
as for QCD at $\mu = 0$. While our focus is clearly on features in a limited 
temperature range at $T \geq T_c$, also some section of the low-temperature 
region can be successfully accomodated in the model, leaving the deep IR region 
for further studies. We also stress that we do not require a specified behavior of the 
model outcome in the UV region. 
Note here that the influence of both asymptotic regimes
on the equation of state in the considered temperature interval 
is fairly small: as shown in \cite{RYBK} the influence of the UV region 
on dimensionless thermodynamic quantities should not exceed a few percent; 
for $T > 0.7 \, T_c$, the deep IR region 
contributes to $p$ and $I$ as a small integration constant, while  
$s$, $T$, $\vsq$, and the viscosities are independent of it. 
Hence, although our dilaton potentials ignore QCD 
features at $T \rightarrow 0$ and $T \rightarrow \infty$, we
argue that they qualify for further investigations. For instance, supplemented by 
a fit of the quark number susceptibility one can repeat the analysis of \cite{Gubser_quarks}  
with an up-to-date input to a holographic study of the phase diagram. 
Even prior to that we note the interesting drop of the ratio $\zeta/\eta$ by 
$50 \%$ at $T_c$ when including quarks. 

Another obvious extension of our studies would be the inclusion of a field dual to 
the chiral condensate $\langle q \bar q \rangle$, which is responsible not only for the breaking of
conformal invariance in addition to the gluon condensate as expressed by the trace anomaly 
but, even more importantly, it is in the chiral limit an order parameter of 
chiral symmetry breaking in QCD. Extensive investigations in this direction, 
albeit for the Veneziano limit QCD, were performed in \cite{Kiritsis_VQCD}, where 
chiral symmetry breaking is realized by tachyon dynamics.

In summary, we present an adjustment of a single-field dilaton potential 
to recently confirmed lattice QCD thermodynamics data in the temperature range $(0.7 - 3.5)T_c$. 
A criterion is delivered for ensuring a cross-over at the softest point.

Acknowledgements:
The work is supported by BMBF grant 05P12CRGH1 and European Network HP3-PR1-TURHIC.
Inspiring discussions with J.\ Randrup are gratefully acknowledged.

\end{document}